\documentclass[10pt,conference,a4paper]{IEEEtran}
\usepackage{times,amsmath,epsfig}
\title{Performance Enhancement of Multiuser Time Reversal  UWB Communication System}
\author{%
{I. H. Naqvi,   A. Khaleghi,   G. Elzein }%
\vspace{1.6mm}\\
\fontsize{10}{10}\selectfont\itshape
Institute of Electronics and Telecommunication Rennes, IETR-UMR CNRS 6164\\
INSA, 20 Avenue des Buttes de Coesmes ,  35043, Rennes France.\\
\fontsize{9}{9}\selectfont\ttfamily\upshape
$~^{1}$Ijaz-Haider.Naqvi@ens.insa-rennes.fr\\
$~^{2}$Ali.Khaleghi@insa-rennes.fr%
$~^{3}$Ghais.El-Zein@insa-rennes.fr %
\vspace{1.2mm}\\
\fontsize{10}{10}\selectfont\rmfamily\itshape
}
\begin{document}
\maketitle

\begin{abstract} 
UWB communication is a recent research area for indoor propagation channels. Time Reversal (TR) communication in UWB has shown promising results for improving the system performance. In multiuser environment, the system performance is significantly degraded due to the interference among different users. TR  reduces the interference caused by multiusers due to its spatial focusing property. The performance of a multiuser TR communication system is further improved if the TR filter is modified. 
In this paper, multiuser TR in UWB communication is investigated using simple TR filter and a modified TR filter with circular shift operation. The concept of circular shift in TR is analytically studied. Thereafter,  the channel impulse responses (CIR) of a typical indoor laboratory environment are measured. The measured CIRs are used to analyze the received signal peak power and signal to interference  ratio (SIR) with and without performing the circular shift operation in a multiuser environment. %Multiuser communication in UWB with Time Reversal (TR) is novel and promising research area.  Improvement in the system performance is one of the goals of UWB communication. 
%If instead of using the time reversed version of the channel impulse response as a transmitter prefilter,  a circularly shifted time reversed channel impulse response is used, it can result in a better system performance. In the context of Time Reversal (TR) communication,  circular shift can be quite significant. In a  multi-user scenario, it reduces the interference significantly. This results in an improved performance as compared to the simple TR (without circular shift). But  circular shift has its limitations and these limitations must be considered. In this paper, we have analyzed the effects of circularly shifted Time Reversal on Received Peak Power and Signal to Interference Ratio (SIR).
\end{abstract}

% NOTE keywords are not used for conference papers so do not populate them
\section{Introduction}
\label{intro}
% no \IEEEPARstart
Ultra Wide Band is been thought as a band in which high data rate communication can take place for short range applications. In particular, research is aimed for allowing communication with low complexity devices.  The use of Rake receivers for signal detection has already been studied for communication in dense multipath environment [2],[3]. TR can reduce the effects of inter symbol interference (ISI) significantly and can eliminate the need of high complex equalizers at the receiver \cite{basic}. 
TR can be exploited as a communication scheme with very simple receivers. Classically, TR has been applied in acoustics and under water communication applications [4]-[6]. In [5] error free communication is demonstrated in the ultrasonic frequency regime and in a multiuser scenario. It is because of these characteristics that TR is gaining more and more attention for communication in UWB. Recent research has shown the potential of TR in wireless communication and specially in UWB short range communication [6]-[8]. Techniques have been proposed for utilizing TR with Minimum Mean Square Equalizer (MMSE) in UWB [9]. 

In \cite{basic} TR is applied to multiusers in order to improve the multiuser system capacity and communication range. UWB TR helps in improving temporal compression and spatial focusing. To improve further the performance in a multiuser scenario, authors in \cite{basic} have proposed to apply the circular shift (CS) operation on the transmitter pre-filter to reduce the effects of interference caused by other users. There are some differences in our approach to the one used in \cite{basic}. First, only one transmitting antenna is used instead of a multi element antenna.
%as the presence of multi elements only improve the received signal power but the overall shape of the curves remain same.  
Secondly,  the system performance by varying the amount of CS  is studied, also  the SIR for different users is independently evaluated.

The rest of the paper is organized as follows. First, a review of TR is presented in section \ref{tr}. The effect of CS on TR is analyzed in section \ref{cstr}.  Experimental measurement setup and  results  are presented and analyzed in Section \ref{results}.  Finally, Section \ref{conclusion}  concludes this paper. 

\section{Time Reversal Review }

\label{tr}
TR is essentially a pre-Rake scheme in which time reversed channel impulse responses (CIR) are used as transmitter pre-filter. The signal (after being pre-filtered) propagates in an invariant channel following the same paths and results in coherently adding all the received signals in the delay and spatial domains. 
With this technique, a focusing gain in the order of 8dB for indoor propagation channel, strong temporal compression and spatial focusing (depending on the signal band width) are observed [10]. The received signal quality is improved by the focusing gain, ISI effects are mitigated by temporal compression  and multiuser interference is reduced due to spatial focusing. The received signal at the intended receiver $ (j) $ can be mathematically represented as:
\begin{equation}
s(t)\star h_{ij}(-t)^{\ast}\star h_{ij}(t) = s(t)\star R_{ij}^{auto}(t)
\label{eq1}
\end{equation}
where $h_{ij}(t) $  is the CIR from the transmitting point to an intended receiver, $ s(t)$ is the transmitted signal, $ \star $ denotes convolution and $ (.)^{\ast} $ means the complex conjugate of the function and $  R_{ij}^{auto}(t)$ is the autocorrelation of the CIR between the  transmitting antenna $ (i) $ and  receiving antenna $ (j)$.
The received signal at any non intended receiver $ (k) $ is:
\begin{equation}
s(t)\star h_{ij}(-t)^{\ast}\star h_{ik}(t) = s(t)\star R_{ikj}^{cross}(t)
\label{eq2}
\end{equation}
where $h_{ik}(t) $  is the CIR from the transmitting point to an unintended receiver and $ R_{ikj}^{cross}(t) $ is the cross-correlation of the CIR  $h_{ik}(t) $ and the time reversed complex conjugated version of the transmitted signal $ h_{ij}(-t)^{\ast}$.
If the channels are uncorrelated, then the signal transmitted for one receiver will act as a noise for a receiver at any other location. This means that the channel itself codes the transmitted signal orthogonally and results in a secure communication with low probability of detect and low probability of intercept. If there are $N_{r} $ receiving antennas and one transmitting antenna, the received signal by the  $ jth$  receiving antenna is:
\begin{equation}
y_{j}(t) = \underbrace{s_{j}(t) \star  R_{ij} ^{auto} (t)}_{Signal(j)}   +\underbrace{\sum _{k= 1;k\neq j}^{N_{r}} s_{k}(t)\star R_{ikj}^{cross}(t)}_{Interference(j)}  %\linebreak[1] + \underbrace{n-{j}(t)}_{Noise(j)}
\label{eq3}
\end{equation}
\begin{displaymath}
+ \underbrace{n_{j}(t)}_{Noise(j)}
\end{displaymath}
where $s_{j}(t)$ and $ s_{k}(t)$ are the transmitted signals intended for the $ jth $ user and the $ kth$ user respectively. 
The formula written in (\ref{eq3}) is for  $ N_{r}$ simultaneous SISO systems. If the channels are uncorrelated, the Interference part in (\ref{eq3}) will be negligible, enabling an interference free communication with different users. The SIR for the $ jth$ user can be calculated as:
\begin{equation}
SIR_{j} = 10log_{10} \frac{\arrowvert Signal_{j}(t = t_{peak}) \arrowvert ^{2}}{\arrowvert Interference_{j}(t = t_{peak}) \arrowvert ^{2}}
\label{eq4}
\end{equation}
where $ t_{peak } $ is the decision time or the time at which the $ Signal $  peak is received.
In a multiuser communication scenario, the interference at the peak of the $ Signal $  increases linearly with the number of simultaneous transmissions \cite{basic}. In simple TR scheme, the transmitted power of each symbol for each transmission link is equal to the power of the estimated CIR amplified to a constant. It can be problematic when there is a simultaneous transmission for different communication links. As the intended transmitted power for different receiving antennas can be different, the interference power can be a large fraction of the intended signal. The solution is equal power control as described in \cite{basic}. The time reversed CIR is normalized with the measured wide band power so that the intended power for each receiving antenna is equal:
\begin{equation}
h_{ij}^{TR}(t) = \frac{h_{measured_{ij}}(-t)^{\ast}}{\Arrowvert h_{measured_{ij}}(t)\Arrowvert}
\label{eq5}
\end{equation}
where $ \Arrowvert . \Arrowvert $ denotes the Frobenius norm operation.

In addition to this normalization, the signal must be normalized, so that the total transmitted power always has a fixed unit value. For example, in case of two simultaneous users $ j $ and $ k $, the transmitted signal would be:
\begin{equation}
S_{tx}^{TR} = \frac{h_{ij}^{TR}(t)+h_{ik}^{TR}(t)}{\Arrowvert h_{ij}^{TR}(t)+h_{ik}^{TR}(t) \Arrowvert}
\label{eq6}
\end{equation}
If we want to compare the performance of different users in a multiuser scenario, this normalization must be done, otherwise the total transmitted power for different CIRs will not be equal. 
 
\section{Circular Shift Time Reversal (CSTR)}
\label{cstr}
\begin{figure}[!t]
\centerline{\psfig{figure=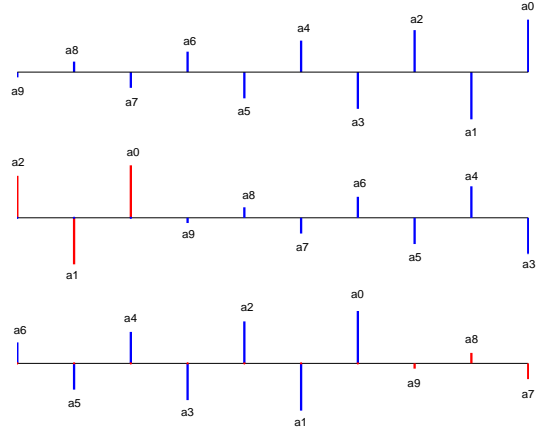,height=64.54mm} }
\caption{a) Time Reversed CIR without CS  b) with right CS of 3 taps  c) with left CS of 3 taps }
\label{lrcs}
\end{figure}
If a circularly shifted time reversed CIR is used as a transmitter pre-filter, the resulting scheme can be called as Circular Shift Time Reversal (CSTR). CSTR can be further classified into left CSTR and right CSTR depending upon the direction of the circular shift operation applied on the time reversed version of CIR.  To elaborate the phenomenon of CSTR, consider a CIR, $ h(t) $,  with $ N $ multipath taps, represented as:
\begin{equation}
h(t) = \sum_{i=1}^{N}\alpha_{i} \delta(t-\tau_{i}) 
\label{eq7}
\end{equation}
 
The time reversed version of the CIR is written as:
\begin{equation}
h(-t) = \sum_{i=1}^{N}\beta_{i} \delta(t-\tau_{i}) 
\label{eq8}
\end{equation}
where $ \beta_{i} = \alpha_{N-i+1} $ and $ \alpha_{i} $ are the coefficients of the CIR $ h(t) $ and  $ \tau_{i} $ is the delay associated with the $ i_{th} $  tap.

If $ h(-t) $ is a discrete time reversed CIR with $ N $  taps that is circularly shifted to right by $ l $ taps, the resulting function is represented as:

\begin{displaymath}
circshift(h(-t),l(right)) =
\end{displaymath}
\begin{equation}
\sum_{i=1}^{l}\beta_{N-l+i} \delta(t-\tau_{i}) +\sum_{i= l+1}^{N} \beta_{i-l} \delta (t-\tau_{i})
\label{eq9}
\end{equation}

Fig. \ref{lrcs} shows a time reversed CIR without any circular shift, with right CS of 3 taps and with left CS of 3 taps. The equation for the  left circular shift can be written as:
\begin{displaymath}
circshift(h(-t),l(left)) = 
\end{displaymath}
\begin{equation}
\sum_{i=1}^{N-l}\beta_{i+l} \delta(t-\tau_{i}) +\sum_{i= N-l+1}^{N} \beta_{i+l-N} \delta (t-\tau_{i})
\label{eq10}
\end{equation}

where $ l $ can have values:
\begin{displaymath}
 1\leq l \leq N-1
\end{displaymath}

As described in the previous section, the interference at the peak increases with the number of simultaneous transmissions. Simultaneous transmission with the simple TR scheme means that the maximal peaks of the time reversed impulse responses are aligned with each other in order to be transmitted. This results in the creation of the interference which is equal to the magnitude of the CIRs cross-correlations ($ R_{ijk}^{cross} $). This interference will increase to the sum of  $ N -1  $ cross-correlations for $ N $ simultaneous users.  If CSTR is used instead of simple TR, the effects of the interference can be greatly reduced. If symbols for more than one user are transmitted simultaneously with CSTR, the maximal peaks of the time reversed impulse responses will no longer be aligned with each other. The taps in the propagating channel containing more energy are multiplied by the taps of the transmitted signal with less energy and vice versa. This results in greatly reduced interference with CSTR.

Received signal for \emph{right CSTR} of $ l $ taps and $  s(t) = 1 $ is written as:
\begin{equation}
R_{x}^{CSTR} = circshift(h(-t),l(right))^{\ast}\star h(t) 
\label{eq11}
\end{equation}
 Applying (\ref{eq9}), the equation becomes: 
%\begin{displaymath}
%R_{x}^{CSTR} = (\sum_{i=1}^{l} \alpha_{N-l+i} \delta(t-\tau_{i}) + \sum_{i= l+1}^{N} \alpha_{i-l} \delta(t-\tau_{i})) 
%\end{displaymath}
%\begin{displaymath}
%\star  \sum_{i=1}^{N} a_{i} \delta(t-\tau_{i}) 
%\end{displaymath}
\begin{displaymath}
R_{x}^{CSTR} =   \Big( \underbrace{\sum_{i=1}^{l} \beta_{N-l+i} \delta(t-\tau_{i}) \star \sum_{i=1}^{N} \alpha_{i}  \delta(t-\tau_{i})}_{Image} \Big )+ 
\end{displaymath}
\begin{equation}
 \Big( \underbrace{\sum_{i= l+1}^{N} \beta_{i-l} \delta(t-\tau_{i}))  \star \sum_{i=1}^{N} \alpha_{i}  \delta(t-\tau_{i})}_{Signal}\Big)
\label{eq12}
\end{equation}

%\begin{equation}
%circshift(h(n),l) = \left\{ \begin{array}{ll} 
%h(n+l) & \textrm{for $n \leq(L-l)$}\\
%h(n-(L-l)) & \textrm{for $n>(L-l)$}
%\end{array} \right.
%\label{eq7}
%\end{equation}

The received signal for a  right CSTR  consists of two parts. First part is the convolution of the first $ l $ taps of  the  right CSTR  transmitted signal (i.e. the last $ l $ taps of the time reversed CIR moved  to the start) with the CIR and the second part is the convolution of the rest of the taps with CIR. These two convolutions result in two signals with two significant  peaks. The later is called $ Signal $ and the former as $ Image$. The position of the $ Signal $ peak also moves from center toward right by $ l $ taps. The distance between the position of the $ Signal $ and $ Image $  peaks is  always equal to $ N $. As the amount of the circular shift increases so does the amplitude of the $ Image $ peak depending upon the power contained in the shifted taps. If the power of shifted taps (i.e. first $ l $ taps) becomes greater than the power of the rest of the taps, the amplitude of the $ Image $ peak becomes greater than the $ Signal $ peak. Thus, the amplitude of the $ Signal $ and $ Image $ peaks depends directly on the power of the taps responsible for their creation. 
To summarize, the first $ l $ terms of the circularly shifted time reversed CIR are responsible for the creation of $ Image $ in case of  right CSTR. In case of  left CSTR, the last $ l $ terms of the circularly shifted time reversed CIR are responsible for the creation of $ Image $. 

 Thus, the reduction in the interference with  CSTR  has its cost. On one hand, CSTR improves the performance of the system by reducing the interference caused by the simultaneous transmissions, but on the other hand, it results in the degradation of the system performance as the received signal peak amplitude is reduced with the increase of circular shift.  This behavior of the CSTR was not elaborated by the authors in \cite{basic}. In this paper, we have shown that CSTR has its limits and we can get a better performance with only if these are taken into account. Otherwise, CSTR could lead to a very poor performance.

\section{Experimental Setup and Simulation Results}
\label{results}
	Experiments are performed in a typical indoor environment.  The environment is an office space of  $ 14m \times 8m $.  The frequency responses are measured using vector network analyzer (VNA) in the frequency range of 0.7-2 GHz and with a frequency resolution of 2.24MHz. Two wideband conical mono-pole antennas are used in cross-polar form for the transmitter-receiver link. The height of the transmit antenna is 1.5m and that of receive antenna is 1.6m from the floor. The receiver antenna is moved over a rectangular surface ($ 65cm \times 40cm$) with a precise  positioner system. The frequency responses between the transmitting antenna and receiving virtual array of 35 positions  are measured. The measurements are corrected to compensate the loss of the cables.  The time domain CIRs are computed using the IFFT transformation of the measured frequency responses.  

 %A $ ns $ pulse signal (which is much smaller than rms channel delay spread) is transmitted from one transmitting antenna toward a receiving antenna which is moved over a rectangular surface with a horizontal resolution of 5cm and vertical resolution of 10cm with the help of a precise positioner. 243 Channel Impulse Responses (CIR) are measured in the range of 0.7-2GHz band using Vector Network Analyzer (VNA). Out of these 243, two sets 35 CIRs are selected for the experiments with different interference.  With these 35 CIRs selected for experiments we made $ 1190 (35 \times 34)$  and $ 1085(35 \times 31)$ combinations for simulating two and five users respectively.
% $ 1155 (35 \times 33) $combinations for simulating 3 users, $1120(35 \times 32) $combinations for simulating 4 users and finally $ 1085(35 \times 31)$ combinations for simulating 5 users 
\subsection{Effects of CSTR on Received Signal Peak} 
\label{p_red_cstr}

\begin{figure}[!t]
\centerline{\psfig{figure=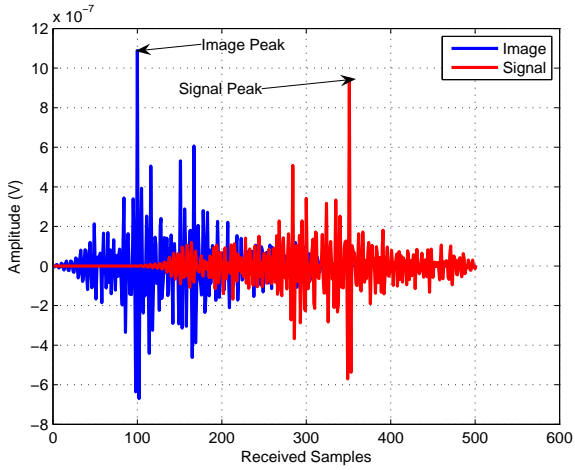,height=64.54mm} }
\caption{Received $ Signal $  and $ Image $ with 40\% right circular}
\label{signal_image}
 \end{figure}

In \cite{basic} it is shown that Interference between  users is greatly reduced with CSTR, but the authors have not studied the effects of power loss due to the CS. We have found that the received signal peak  is reduced with CSTR depending upon the amount of shift. The lost power appears as an $ Image $ and is thus lost. These $ Image $ and $ Signal $ are represented mathematically in (\ref{eq12}). The amplitude of the $ Image $ peak  becomes comparable to the amplitude of the $ Signal $ peak when the power of the shifted taps approaches the power of rest of the taps.  Fig. \ref{signal_image} shows  the received $ Signal $  and  $ Image $  for 40\% right circular shift. 

%\begin{figure}[!t]
%\centerline{\psfig{figure=signal_image_and_difference,height=60.54mm} }
%\includegraphics[scale=0.5]{signal_image_and_difference} 
%\caption{Received $ Signal $ and the sum of $ Siganal $ and $ Image $ with 40\% right circular}
%\label{signal_image_and_difference}
 %\end{figure}
As shown the $ Image $ peak is stronger than the $ Signal $ peak for 40\% right CS bacause the power of the shifted taps at 40\% right CS is more than the power of the rest of the taps.  
When the amplitudes of $Signal $ peak  and $ Image $ peak are added,  it becomes equal to the amplitude of the received signal without any circular shift. 
%The sum of the $ Signal $  and $ Image $ (i.e. RHS of the (\ref{eq12})), the received signal (i.e. LHS of the (\ref{eq12})) and the difference of the two are shown in Fig. \ref{signal_image_and_difference}.  We can see that the former two completely overlaps one another. The fact is further ascertained by the difference of the two being zero. Thus our mathematical analysis of CSTR is confirmed by our experimental results. It must be noted that these curves are for single user, so there is no interference present.

\begin{figure}[!t]
\centerline{\psfig{figure=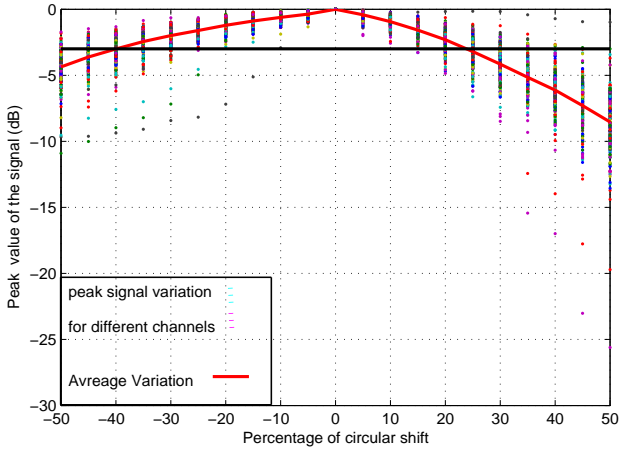,height=64.54mm} }
\caption{Variation of the received signal peak values with different values of the circular shift}
\label{pk_sig_variation}
\end{figure}

Fig. \ref{pk_sig_variation} shows the variation of the $ Signal $ peak power (normalized to the received peak power without CS) with the implied CS for 35 measured channel scenarios. The average variation is also illustrated. As shown the received $ Signal $ peak power reduces as the percentage of the CS (either left or right) increases. Obviously, the figure can not be symmetric for left and right circular shift because the amount of power reduced due to the CS is directly related to the power of the shifted taps.  The power of the shifted taps for  right CSTR  is more than the power of the shifted taps for  left CSTR  as for  right CSTR, the maximal power paths circulate and come at the start of the signal (see Fig. \ref{lrcs}b), whereas in case of  left CSTR, these are the minimal power taps which moves to the end after the CS (see Fig. \ref{lrcs}c). Thus the reduction in the received $ Signal $ peak power with right CSTR is much more than reduction with left CSTR.

\begin{figure}[!t]
\centerline{\psfig{figure=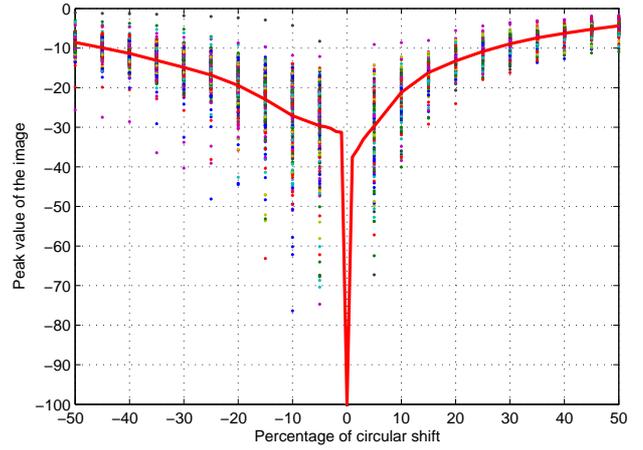,height=64.54mm} }
\caption{Variation of the principal image peak values with different values of the circular shift}
\label{pk_image_variation}
\end{figure}

The variation of the $ Image $ peak with different values of the CS is opposite to the variation of $ Signal $ with the CS i.e. the $ Image $ peak increases with the CS (see Fig. \ref{pk_image_variation}).
When  the sum of $ Signal $ and $ Image $ peak amplitudes (not power)  is taken, it is comes out to be exactly equal to the peak amplitude of the received signal peak without CS ascertaining that the received signal peak is decomposed into a $ Signal $ peak and an $ Image $ peak after CS. 

\subsection{Effects of Circular Shift on Signal to Interference Ratio (SIR)}
\label{SIR}

When symbols for more than one user are transmitted simultaneously, only one part amongst $ N $ simultaneously transmitted signal is intended toward a specific user. The rest $ N-1 $  parts are the Interferences for that user. In multiuser CSTR the intended signals for different users are circularly shifted with different CS percentages. In \cite{basic} the CS percentage increases by 5\% for every user. The authours have shown an improvement in the SIR curves with the CS. Maximum shift percentage studied in \cite{basic} is 20\% for five simultaneous users. For more users the percentage of CS  must be increased. We have also considered a scenario of five simultaneous users but we have gradually increased the CS percentage for User5 till it reaches 30\% (in both directions). The purpose is to study the effects of CS on SIR.  The intended signals for five users are transmitted simultaneously with a CS difference of 3\%. Thus  every user receives four intrerferences along with its intended signal. We have varied (left and right) CS percentage for User5 from 12-30\%. The CS percentages for other four users remain same (i.e. 0-15\%).

In the context of multiuser CSTR, it is important  to study the system performance on user to user basis otherwise the results might not show the true picture of the scenario. For example in our studied scenario with five simultaneous users, the performance of User1 will always be better than other users. The intended $ Signal $ for User1 does not undergo CS while all four of its $ Interference $ parts do undergo CS. The benefits are two fold; first its $ Interference $ is reduced considerably due to the CS and second no power is lost as an $ Image $. Thus,  if the total performance of the system is studied, it is not a fair representation.  That is why, we have presented the curves for User5 which is worst case i.e. its intended $ Signal $ undergoes largest CS while  its $ Interference $ undergoes least CS.

\begin{figure}[!t]
\centerline{\psfig{figure=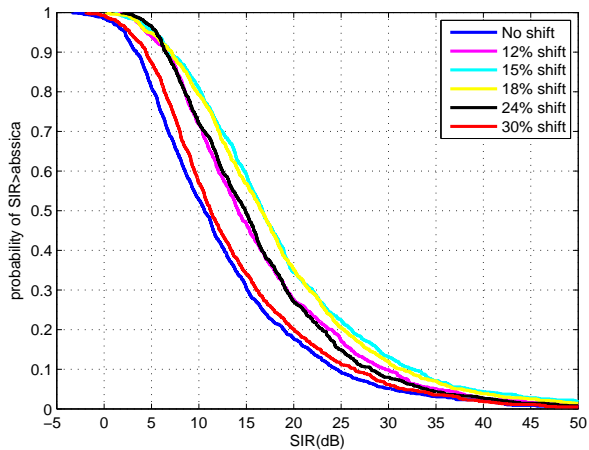,height=64.54mm} }
\caption{CDF of SIR with right  circular shift for User5 with different shift \% }
\label{cdf_sir_rcs_u5} 
\end{figure}
\begin{figure}[!t]
\centerline{\psfig{figure=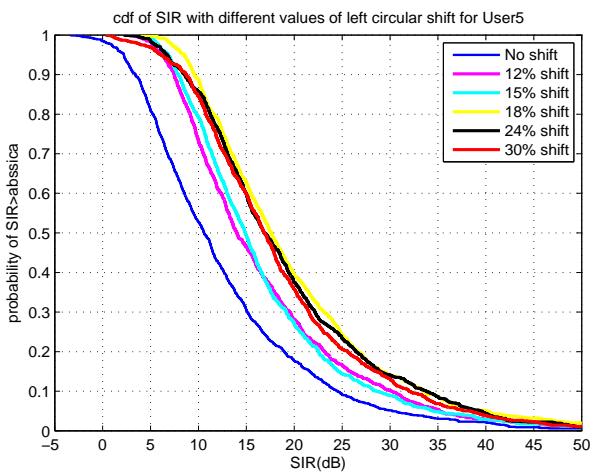,height=64.54mm} }
\caption{CDF of SIR with left circular shift for User5 with different shift \%}
\label{cdf_sir_lcs_u5} 
\end{figure}

%For a detailed analysis of the effects of CSTR on SIR, we simulated for the cumulative density function (cdf) of SIR from 50\% left circular shift to 50\% right circular shift i.e. from -50\% to +50\% for five simultaneous users. 
The CSTR (left or right) scheme results in  the reduction of the interference caused by the other users. 
%The reasons for this reduction has already been explained (see Section \ref{CSTR}). 
This reduction in  the interference  results in an increase of SIR. To simulate the scenario of 5 simultaneous users, we have generated $ 1085(35 \times 31)$ combinations of 5  from the existing 35 measured CIRs.
Figs. \ref{cdf_sir_rcs_u5}-\ref{cdf_sir_lcs_u5} show the curves of cdf of SIR for User5 for different (left and right) CS. As shown the performance is improved with CSTR. For right CSTR (Fig. \ref{cdf_sir_rcs_u5}),  the improvement in SIR reaches its saturation state, and the performance degradation starts after certain value of CS (20\%). This degradation in the performance after certain value of CS is due the reduction in received signal strength with CS. However, for left the performance degradation is lesser as the reduction in the received signal strength is considerably lesser for left CSTR than right CSTR (see Fig. \ref{cdf_sir_lcs_u5}).

\section*{Acknowledgment}

This work was partially supported by ANR Project MIRTEC and French Ministry of Research.

\section{Conclusion}
\label{conclusion}
In this paper, TR is used with circular shift (i.e. CSTR) and is compared with the simple TR scheme. It is shown that in a multiuser scenario, CSTR reduces the interference considerably and results in a better SIR performance than simple TR scheme. The SIR improvement reaches its limit and the performance starts degrading due to the reduction in received peak with the CS.  Thus, CSTR should be used with care as it can result in a very poor performance for some users with high values of circular shift specially in the case of right CSTR. It is also shown that the reduction in the received peak power depends directly on the power of the shifted taps and for left CSTR the reduction is considerably lower than right CSTR.

%\bibliographystyle{IEEEtran}

%\bibliography{IEEEabrv,IEEEexample}

\end{document}